\documentclass[12pt,onecolumn]{article}
\usepackage{epsf}
\usepackage[dvips]{graphics}
\title{
Conditions for the emergence of spatial asymmetric retrieval states in
attractor neural network}
\author{
Kostadin Koroutchev \footnote{
Corresponding author; k.koroutchev@uam.es;
Inst. for Computer Systems, Bulgarian Academy of Sciences, 1113, Sofia, Bulgaria}
$^{(1)}$ and Elka Korutcheva \footnote{
G.Nadjakov Institute of Solid State Physics, Bulgarian Academy of Sciences, 1784, Sofia, Bulgaria}$^{\ (2)}$
}
\begin{document}
\maketitle
{\center\small
$^{(1)}$ Depto. de Ingenier\'{\i}a Inform\'{a}tica\\
Universidad Aut\'{o}noma de Madrid, 28049 Madrid, Spain\\
$^{(2)}$ Depto. de F\'{\i}sica Fundamental,\\
Universidad Nacional de Educaci\'on a Distancia,\\
c/Senda del Rey, No 9, 28080 Madrid, Spain\\
}

\abstract{
In this paper we show that during the retrieval process in a binary symmetric Hebb neural network, spatial localized states can be observed when the connectivity of the network is distance-dependent and constraint on the activity of the network is imposed, which forces different level of activity in the retrieval and learning states. 
This asymmetry in the activity during the
retrieval and learning is found to be sufficient condition 
in order to observe spatial localized retrieval states.
The result is confirmed analytically and by simulation.
}
{{\bf Keywords:} neural networks, spatial localized states, replica formalism}\\
{\bf PACS:} 64.60.Cn, 84.35.+i, 89.75.-k, 89.75.Fb

\section{Introduction}

In a recent publication \cite{AleYasser} it was shown that using
linear-threshold  model
neurons, Hebb learning rule,
sparse coding and distance-dependent asymmetric connectivity,
spatial asymmetric retrieval states (SAS) can be observed. 
This asymmetric states are characterized by a spatial localization of the
activity of the neurons, described by the formation of local bumps.

Similar results have been reported in the case of Hebb binary model for 
associative neural network \cite{EK}. 
The observation is intriguing, because all components of the network
are intrinsically symmetric with respect to the positions of the neurons
and the retrieved state is clearly asymmetric. 
An illustration of SAS in binary network is presented in Fig.\ref{figbulb}.

In parallel to the present investigation, extensive computer simulations 
have been performed in the case of integrate and fire neurons \cite{qbio}, 
where bump formations were also reported. These results are in agreement 
with our previous results \cite{EK} that networks of binary neurons do not 
show bumpy retrieval solutions when the stored and retrieved patterns 
have the same mean activity.

The biological reason for this phenomenon is based on the transient 
synchrony that leads to recruitment of cells into local bumps, followed by 
desynchronized activity within the group \cite{Rubin},\cite{Brunel}.

When the network is sufficiently diluted, say less then 5\%,
then the differences between asymmetric and symmetric connectivity
are minimal \cite{thebook}.
For this reason we expect that
the impact of the asymmetrical connectivity will be minimal and
the asymmetry could not be considered as necessary condition for
the existence of SAS, as it is confirmed in the present work.

There are several factors that possibly contribute to the SAS  
in model network.

In order to introduce the spatial effects in neural networks(NN),
one essentially needs
distance measures and topology between the neurons,
imposing some distribution on the connections,
dependent on that topology and distances.
The major factor to observe spatial asymmetric activity is of
course the spatially dependent connectivity of the network.
Actually this is an essential condition, because
given a network with SAS, by applying random permutation to the enumeration
of the neurons, one will obviously achieve states without SAS. Therefore,
the topology of the connections must depend on the distance
between the neurons.

Due to these arguments, a symmetric and distance-dependent connectivity
for all neurons is chosen in this study.

We consider an attractor NN model of Hebbian type formed  by $N$ binary 
neurons $\{S_i\}, S_i\in \{-1,1\}, i=1,...,N$,
storing $p$ binary patterns $\xi_i^{\mu}, \mu\in \{ 1...p\}$,
and we assume a symmetric connectivity between the neurons 
$c_{ij}=c_{ji}\in\{0,1\}, c_{ii}=0$. $c_{ij}=1$ 
means that neurons $i$ and $j$ are connected.
We regard only connectivities in which the fluctuations between
the individual connectivity are small, e.g.
$\forall_{i} \sum_j c_{ij}\approx c N$,
where $c$ is the mean connectivity.

The learned patterns are drawn from the following distribution:
\[
P(\xi_i^\mu)=\frac{1+a}{2}\delta(\xi_i^\mu-1+a)+\frac{1-a}{2}\delta(\xi_i^\mu+1+a),
\]
where the parameter $a$ is the sparsity of the code.
Note that the notation is a little bit different from the usual one. 
By changing the variables 
$\eta\rightarrow \xi+a$ 
and substituting in the above equation, one obtains the usual form for the
pattern distribution in the case of sparse code \cite{feigelman}.

Further in this article we show that
imposing symmetry between the retrieval and the learning states,
i.e. equal probability distributions of the patterns and
the network activities, no SAS exists. Spatial asymmetry
can be observed only when asymmetry between
the learning and the retrieval states is imposed.

Actually, by using binary network and symmetrically distributed patterns,
the only asymmetry between the retrieval and the learning states
that can be imposed, independent on the position of the
neurons, is the total number of the neurons in a given state. Having in
mind that there are only two possible states, this condition leads to a
condition on the mean activity of the network.

To impose a condition on the mean activity,
we  add an extra term $H_a$ to the Hamiltonian
\[
H_a = N R \sum_i S_i/N.
\]
This term actually favors states with
lower total activity $\sum_i S_i$ that is equivalent to
decrease the number of active neurons, creating
asymmetry between the learning and the retrieval states.
If 
the value of the parameter $R=0$, the corresponding model has been intensively studied since the classical results of Amit et al. \cite{Amit1} for symmetrical code and Tsodyks, Feigel'man \cite{feigelman} for sparse code. 
These results show that the sparsities of the learned patterns and the retrieval states are the same and equal to $aN$.
In the case $R\neq 0$, the energy of the system increases with the number of active neurons ($S_i=1$) and the term $H_a$ tends to limit the number of active neurons below $aN$. 
Roughly speaking, the parameter $R$ can be interpreted as some kind of chemical potential for the system
to force a neuron in a state $S=1$.
Here we would like to mention that Roudi and Treves \cite{AleYasser} 
have also stated that the activity of the linear-threshold network, 
they study, 
has 
to be constrained in order to have spatially asymmetric states. 
However, no explicit analysis was presented in order to explain the phenomenon
and the condition is not shown to be sufficient one, probably because the use
of more complicated linear threshold neurons 
diminishes the importance of this condition.
Here we point out the importance 
of the constraint on the network level of activity and show 
that it is a sufficient condition for observation of spatial
asymmetric states in binary Hebb neural network.

The goal of this article is to find minimal necessary conditions where SAS
can be observed.
We start with general sparse code and sparse
distance-dependent connectivity
and
using replica symmetry paradigm we find the equations for the order
parameters.

Then we study the solutions of these equations using some approximations
and compare the analytical results
with simulation.

The conclusion, drawn in the last part, shows that only the
asymmetry between the learning and the retrieval states is
sufficient to observe SAS.


\section{Analytical analysis}

\subsection{Connectivity matrix}
\begin{figure}[t]
\begin{center}
\epsfxsize 5.7cm \epsfysize=5cm
\epsfbox{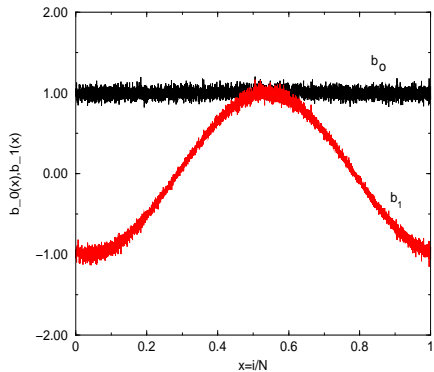}
\caption{\small
The components of the first eigenvectors of the connectivity matrix, normalized
by a square root of their corresponding eigenvalue, in order to eliminate the effect of the
size of the network. The first eigenvector has a constant component, the
second one-sine-like component. $N=6400, c=320/N, \lambda_0=319.8, \lambda_1=285.4$.
}
\label{fig4}
\end{center}
\end{figure}

For the analytical analysis of the SAS states, we
consider the decomposition of the connectivity matrix
$c_{ij}$ by its eigenvectors $a_i^{(k)}$:
\begin{equation}
\label{connectiv}
\nonumber
c_{ij}=\sum_k \lambda_k a_i^{(k)} a_j^{(k)},\  \sum_i a_i^{(k)} a_i^{(l)} =\delta_{kl},
\end{equation}
where $\lambda_k$ are the corresponding (positive) eigenvalues.
The eigenvectors $a_i^{(k)}$ are ordered by its corresponding
eigenvalues in decreasing order, e.g.
\begin{equation}
\label{akorder}
\forall a_j^{(k)},a_j^{(l)}, k>l \Rightarrow \lambda_k\leq \lambda_l.
\end{equation}
For convenience we introduce also the parameters $b_i^k$:
\begin{equation}
\nonumber
b_i^k\equiv a_i^{(k)}\sqrt{\lambda_k/c}.
\end{equation}
To get some intuition of what $a_j^{(k)}$
 look like, we plot in Fig. \ref{fig4}
the first two eigenvectors $a^{0}$ and $a^{1}$.

For a wide variety of connectivities, the first
three eigenvectors are approximately the following:
\begin{equation}
\label{a1}
a_k^{(0)}=\sqrt{{1}/{N}},\ \
a_k^{(1)}=\sqrt{{2}/{N}}\cos(2\pi k/N),\ \
a_k^{(2)}=\sqrt{{2}/{N}}\sin(2\pi k/N).
\end{equation}

Further, the eigenvalue $\lambda_0$ is approximately the mean connectivity of the network per node, that is the mean number 
of connections per neuron $\lambda_0=c N$.

\subsection{Hamiltonian}
Following the classical analysis of Amit et al. \cite{Amit1}, we study binary Hopfield model \cite{Hopfield}
\begin{equation}
H_{int} = - \frac{1}{c N}\sum_{ij\mu} S_i \xi_i^\mu c_{ij} \xi_j^\mu S_j ,
\end{equation}
where the matrix $c_{ij}$ denotes the matrix of connectivity and we have assumed Hebb's rule of learning \cite{Hebb}.

In order to take into account the finite numbers of overlaps that condense macroscopically, we use Bogolyubov's method of quasi averages \cite{Bogol}.
To this aim we introduce an external field, conjugate to a final number of patterns $\{\xi_{i}^\nu\}, \nu=1,2,...,s$ by adding a term
\begin{equation}
H_h = - \sum_{\nu=1}^{s} h^{\nu} \sum_i \xi_{i}^{\nu} S_i 
\end{equation}
to the Hamiltonian. 

Finally, as we already mentioned in the Introduction, in order to impose some asymmetry in the neural network's states, we also add the term
\begin{equation}
\label{over}
H_a= N R \left(\frac{1}{N} \sum_i S_i\right) \approx N R \overline{S_i b_i^0},
\end{equation}
where we used $b_i^0\approx 1$. 
The equality is exact for equal degree network. 
The over line in eq.(\ref{over}) means spatial averaging: 
$\overline{(._i)} = \frac{1}{N}\sum_{i} (.)$.

The whole Hamiltonian we are studying now is $H = H_{int} + H_h + H_a$:
\begin{equation}
\label{Hamiltonian}
H = -\frac{1}{c N}\sum_{ij\mu} S_i \xi_i^\mu c_{ij} \xi_j^\mu S_j -
    \sum_{\nu=1}^{s} h^{\nu} \sum_i \xi_{i}^{\nu} S_i+
    N R \overline{S_i b_i^0}.
\end{equation}

\subsection{RS solution}

By using the ``replica formalism'' \cite{mezard}, for the averaged free energy per neuron we get:
\begin{equation}
\label{fenergy}
f=\lim_{n\rightarrow 0} \lim_{N\rightarrow \infty} \frac{-1}{\beta n N} ({\langle\langle}Z^{n}{\rangle\rangle} -1),
\end{equation}
where ${\langle\langle}...{\rangle\rangle}$ stands for the average over the pattern distribution $P(\xi_{i}^{\mu})$, $n$ is the number of the replicas, which are later taken to zero and $\beta$ is the inverse temperature.

Following \cite{Amit1}, \cite{Gardner}, the replicated partition function is
represented by decoupling the sites using an expansion of the connectivity matrix $c_{ij}$ over its eigenvalues
$\lambda_{l}, l=1,...,M$ and eigenvectors $a_{i}^{l}$ (eq.\ref{connectiv}):
\begin{eqnarray}
{\langle\langle}Z^n{\rangle\rangle}=
e^{-\beta \alpha N n /2c }
{\biggl\langle\biggl\langle}
 Tr_{S^{\rho}}
  \exp\biggl[
    \frac{\beta}{2 N}
      \sum_{\mu\rho l}\sum_{ij}
            (\xi_{i}^{\mu} S_{i}^{\rho} b_{i}^{l})
            (\xi_{j}^{\mu} S_{j}^{\rho} b_{j}^{l})+ \nonumber\\
      \beta \sum_{\nu} h^{\nu} \sum_{i \rho} \xi_{i}^{\nu} S_{i}^{\rho} -
      \beta R N \sum_{i \rho} b_{i}^{0}S_{i}^{\rho}/N
  \biggl]
{\biggl\rangle\biggl\rangle},
\end{eqnarray}
with $\alpha=p/N$ being the storage capacity.

Following the classical approach of Amit et al.\cite{Amit1}
we introduce variables $m_{\rho k}^{\mu}$ for each replica $\rho$ and each eigenvalue and split the sums over the first $s$ ``condensed''patterns, labeled by the letter $\nu$ and the remaining (infinite) $p-s$, over which an average and a later expansion over their corresponding parameters was done.\footnote{Without loss of generality we can limit ourself to the case of only one pattern with macroscopic overlap $(\nu=1)$. The generalization to $\nu>1$ is straightforward.}
We have supposed that only a finite number of $m_{k}^{\nu}$ are of order one and those are related to the largest eigenvalues $\lambda_k$.

As a next step, we introduce the order parameters (OP)
\[
q_{k}^{\rho, \sigma}
= \overline{(b_{i}^{k})^2 S_{i}^{\rho} S_{i}^{\sigma}},
\]
where $\rho, \sigma = 1,...,n$ label the replica indexes. Note the role of the connectivity matrix on the OP $q_{k}^{\rho, \sigma}$ by the introduction of the parameters $b_{i}^{k}$.

The introduction of the OP $r_{k}^{\rho, \sigma}$, conjugate to
$q_{k}^{\rho, \sigma}$,
and the use of the replica symmetry ansatz \cite{mezard}
$m_{\rho k}^{\nu}=m_{k}^{\nu}, q_k^{\rho, \sigma}= q_{k}$
for
$\rho \neq \sigma$ and
$r_{k}^{\rho, \sigma}= r_{k}$ for
$\rho \neq \sigma$,
followed by a suitable linearization of the quadratic in $S$-terms and the application of the saddle-point method \cite{Amit1}, give the following final form for the free energy per neuron:

\begin{eqnarray}
f& = &\frac{1}{2c}\alpha(1-a^2) + \frac{1}{2} \sum_k (m_{k})^2
    - \frac{\alpha\beta(1-a^2)}{2} \sum_{k} r_k q_{k}
    + \frac{\alpha\beta(1-a^2)}{2} \sum_k \mu_k r_k +\nonumber\\
&+&
\frac{\alpha}{2\beta}\sum_k [\ln(1-\beta(1-a^2) \mu_k + \beta(1-a^2)q_k) -\\
&-&\beta(1-a^2) q_k(1-\beta(1-a^2)\mu_k + \beta(1-a^2) q_k)^{-1}]-\nonumber\\\nonumber\\
&-&
    \frac{1}{\beta}
     \int{\frac{dz e^{-\frac{z^2}{2}}}{\sqrt{2\pi}}}
       \overline{
         \ln 2 \cosh \beta \left(
            z \sqrt{ \alpha (1-a^2) \sum_l r_l b_{i}^{l} b_{i}^{l}}
       + \sum_l m_{l} \xi_i b_{i}^{l}
       + R b_i^0\right)} .\nonumber
\end{eqnarray}
In the last expression we have introduced the variables $\mu_k=\lambda_k/c N$
and we have used the fact that the average over a finite number of patters
$\xi^{\nu}$ can be self-averaged \cite{Amit1}.
In our case however the
self-averaging is more complicated in order to preserve
the spatial dependence of the retrieved pattern.
The detailed analysis will be given in a forthcoming publication \cite{EK1}.

The equations for the OP $r_k$, $m_k$ and $q_k$  are respectively:

\begin{eqnarray}
  r_k=\frac{q_k (1-a^2)}{\left(1-\beta (1-a^2)(\mu_k-q_k)\right)^2} ,
\end{eqnarray}

\begin{eqnarray}
m_k = \int \frac{d z e^{-\frac{z^2}{2}}}{\sqrt{2\pi}}
       \overline{
          \xi_i b_i^k\tanh \beta \left(
            z \sqrt{ \alpha (1-a^2) \sum_l r_l b_{i}^{l}b_{i}^{l} }
       + \sum_{l} m_{l} \xi_i b^l_i
       + R b_i^0\right)}
\end{eqnarray}
and
\begin{eqnarray}
q_k =  \int\frac{dz e^{-\frac{z^2}{2}}}{\sqrt{2\pi}}
       \overline{
          (b_i^k)^2 \tanh^2 \beta \left(
            z \sqrt{\alpha (1-a^2)\sum_{l} r_l b_{i}^{l} b_{i}^{l}}
       + \sum_{l}m_{l} \xi_i b^l_i
       + R b_i^0\right)} .
\end{eqnarray}

At $T=0$, keeping $C_k\equiv\beta (\mu_k-q_k)$ finite and limiting the above system
only to the first two coefficients,
the above equations read:
\begin{eqnarray}
\label{T=0}
  m_0&=& \frac{1-a^2}{4\pi}\int_{-\pi}^{\pi} g(\phi) d\phi\\
  m_1&=& \sqrt{2\mu_1}\frac{1-a^2}{4\pi}\int_{-\pi}^{\pi}g(\phi)\sin\phi\ d\phi\\
  C_0&=& \frac{1}{2\pi}\int_{-\pi}^{\pi} g_c(\phi) d\phi\\
  C_1&=& \frac{\mu_1}{\pi} \int_{-\pi}^{\pi} g_c(\phi) \sin^2\phi\ d\phi\\
  r_k&=& \frac{\mu_k (1-a^2)}{[1-(1-a^2)C_k]^2}\label{tzeron} ,
\end{eqnarray}

where
\begin{eqnarray}
g(\phi)&=&{\rm erf}(x_1)+{\rm erf}(x_2)\\
g_c(\phi)&=&[(1+a)e^{-(x_1)^2}+(1-a)e^{-(x_2)^2}]/[{\sqrt{\pi}y}]\\
x_1&=&{[(1-a)(m_0+m_1\sqrt{2 \mu_1}\sin\phi)+R]}/{y}\\
x_2&=&{[(1+a)(m_0+m_1\sqrt{2 \mu_1}\sin\phi)-R]}/{y}\\
y&=&\sqrt{2\alpha (1-a^2) (r_0+2\mu_1 r_1\sin^2\phi)}\label{tzero1}.
\end{eqnarray}

When we assume $m_1=m_2=...=0, R=0$,
we obtain the result of Gardner \cite{Gardner}.
When additionally $\mu_1=0$ we obtain the result of Amit et al. \cite{Amit1}.
Of course, in this approximation no conclusion about the size of the bump
can be drawn, because the bump characteristic size is fixed to the characteristic size of a sine wave, that is always one half of the size of the network.

\begin{figure}[t]
\begin{center}
\begin{minipage}{5.70cm}
\epsfxsize 5.7cm 
\epsfbox{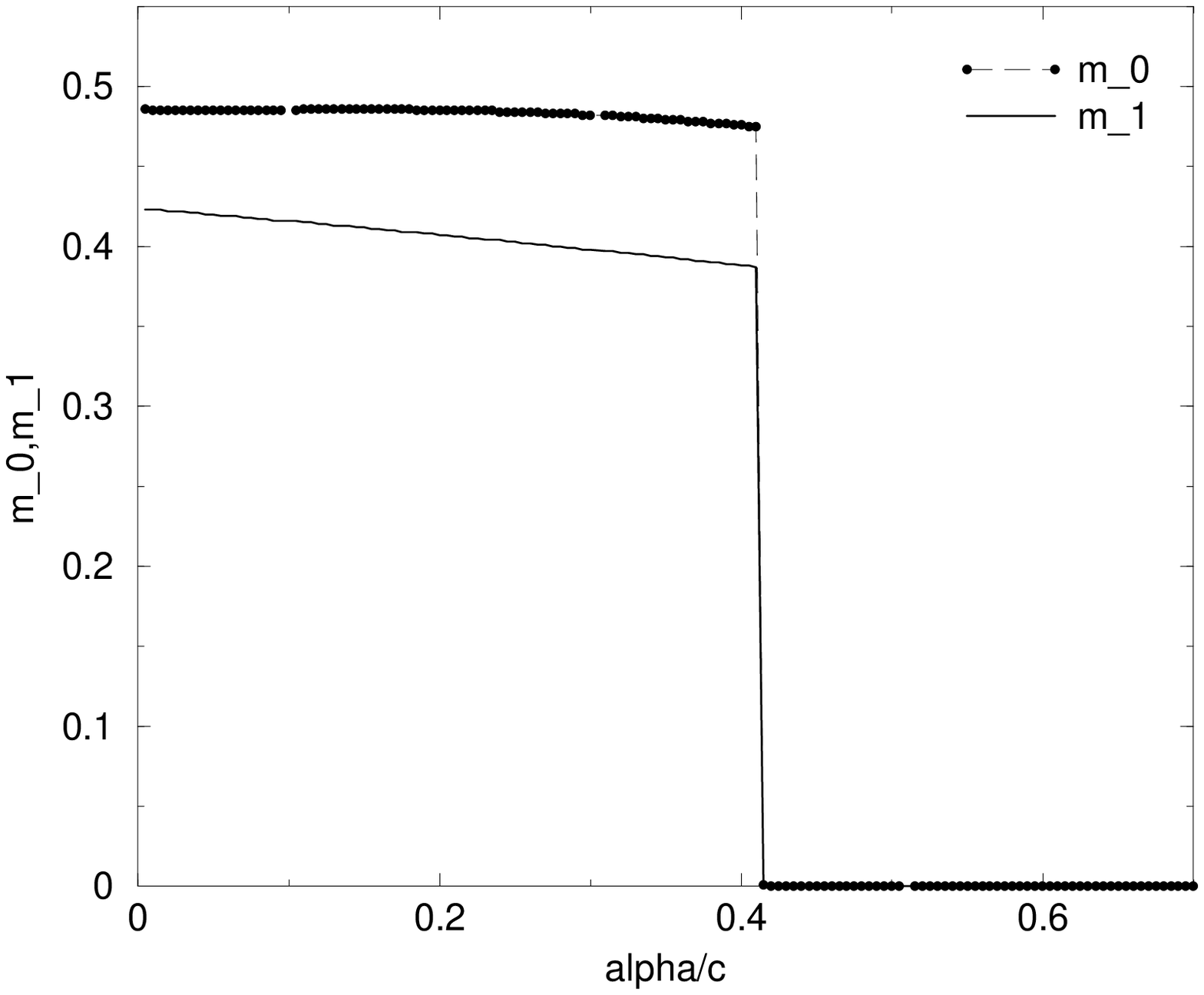}
\end{minipage}
\hfill
\begin{minipage}{5.70cm}
\epsfxsize 5.7cm 
\epsfbox{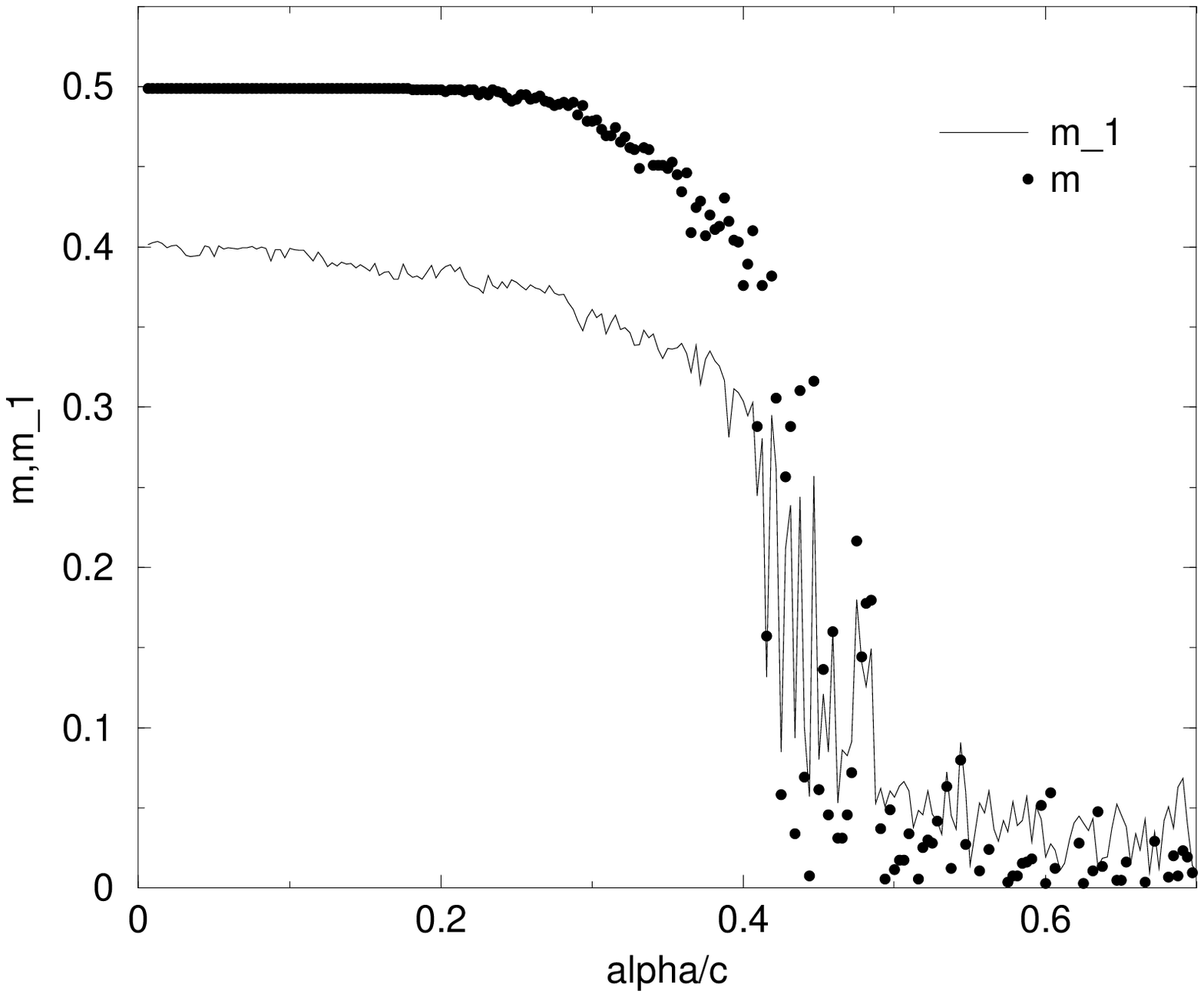}
\end{minipage}
\caption{\small
Left -- Computation of $m_0,m_1$ according to
the eqs.(\ref{T=0}-\ref{tzeron}).
The sparsity of the code is $a=0.2$ and $R=0.57$.
The sharp bound is due to the limitation of the equations
only to the first two terms of the OP $m,r,C$,
as well as to the finite scale effects.
The result of the simulations is represented on the right side.
}
\label{newfig}
\end{center}
\end{figure}

The result of the numerical solution of the eqs.(\ref{T=0}-\ref{tzeron})
is show in Fig. \ref{newfig} (left).
The sharp bound of the phase transition is a result of taking into account
just two terms $k=0,1$ and the lack of finite-size effects in the thermodynamic limit.

\section{Simulations}

In order to compare the results, we also performed computer
simulations. To this aim we chose the network's topology
to be  a circular ring,
with a distance measure
\[
|i-j| \equiv \min(i-j+N\ {\rm mod}\ N,j-i+N\ {\rm mod}\ N)
\]
and used the same connectivity as in Ref.\cite{AleYasser} with typical
connectivity distance
$\sigma_x N$:
\[
P(c_{ij}=1)= c \left[
\frac{1}{\sqrt{2\pi}\sigma_x N} e^{-(|i-j|/N)^2/2\sigma_x^2}+p_0\right].
\]
Here the parameter $p_0$ is chosen to normalize the expression in the brackets.
When $\sigma_x$ is small enough, then spatial asymmetry is expected.

\begin{figure}[t]
\begin{center}
\epsfxsize 5.7cm \epsfysize=5cm
\epsfbox{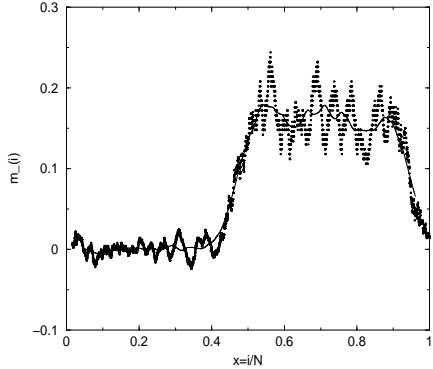}
\caption{\small
Typical bump form. 
The figure shows the running average of the local overlap $m_{(i)}=S_i \xi_i$ vs the position of the neuron (run length 100 and 300).
The sparsity of the code is $a=0.2$,
$R$ is chosen to supply 0.5 of the unconstrained activity. $N=6400, c=0.05$.
}
\label{figbulb}
\end{center}
\end{figure}

In order to
measure the overlap between the states of the patterns
and the neurons and the effect of the single-bump spatial activity during the simulation,
we used the ratio $m_{1}/m_{0}$, where
\[
m_0=\frac{1}{N}\sum_k \xi_k^0 S_k
\]
and
\[
m_{1} = \frac{1}{N} |\sum_{k} \xi_{k}^{0} S_k e^{2\pi i k/N}|.
\]
Because the sine waves appear first,
$m_{1}/m_0$
results also to be a sensitive asymmetric measure,
at least compared to the visual inspection of the running averages
of the local overlap $m_{(i)}\equiv\xi_i S_i$. 
An example of the local overlap form is given in Fig.\ref{figbulb}.

Let us note that $m_{1}$ can be regarded as
the power of the first Fourier component and
$m_0$ can be regarded as a zeroth Fourier component,
that is the power of the direct-current component.

 The corresponding behavior of the order parameters $m_0, m_1$
in the zero-temperature case, obtained by the simulations, is represented in Fig. \ref{newfig} (right).
Note the good correspondence between the numerical
solution of the analytical results and the results obtained by simulation.

\section{Discussion}

\begin{figure}[t]
\begin{center}
\begin{minipage}{5.70cm}
\epsfxsize 5.7cm \epsfysize=5cm
\epsfbox{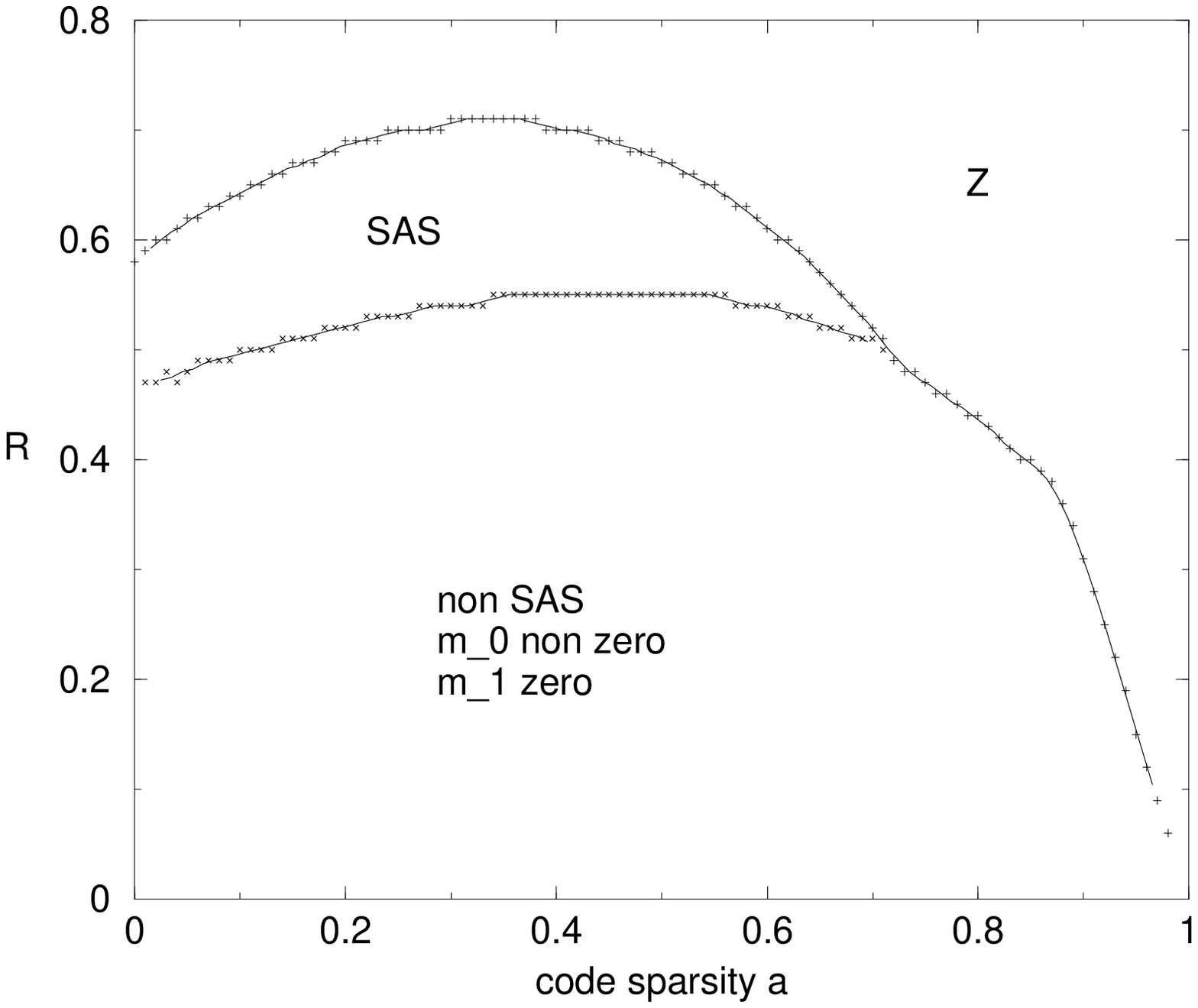}
\end{minipage}
\hfill
\begin{minipage}{5.70cm}
\epsfxsize 5.7cm 
\epsfbox{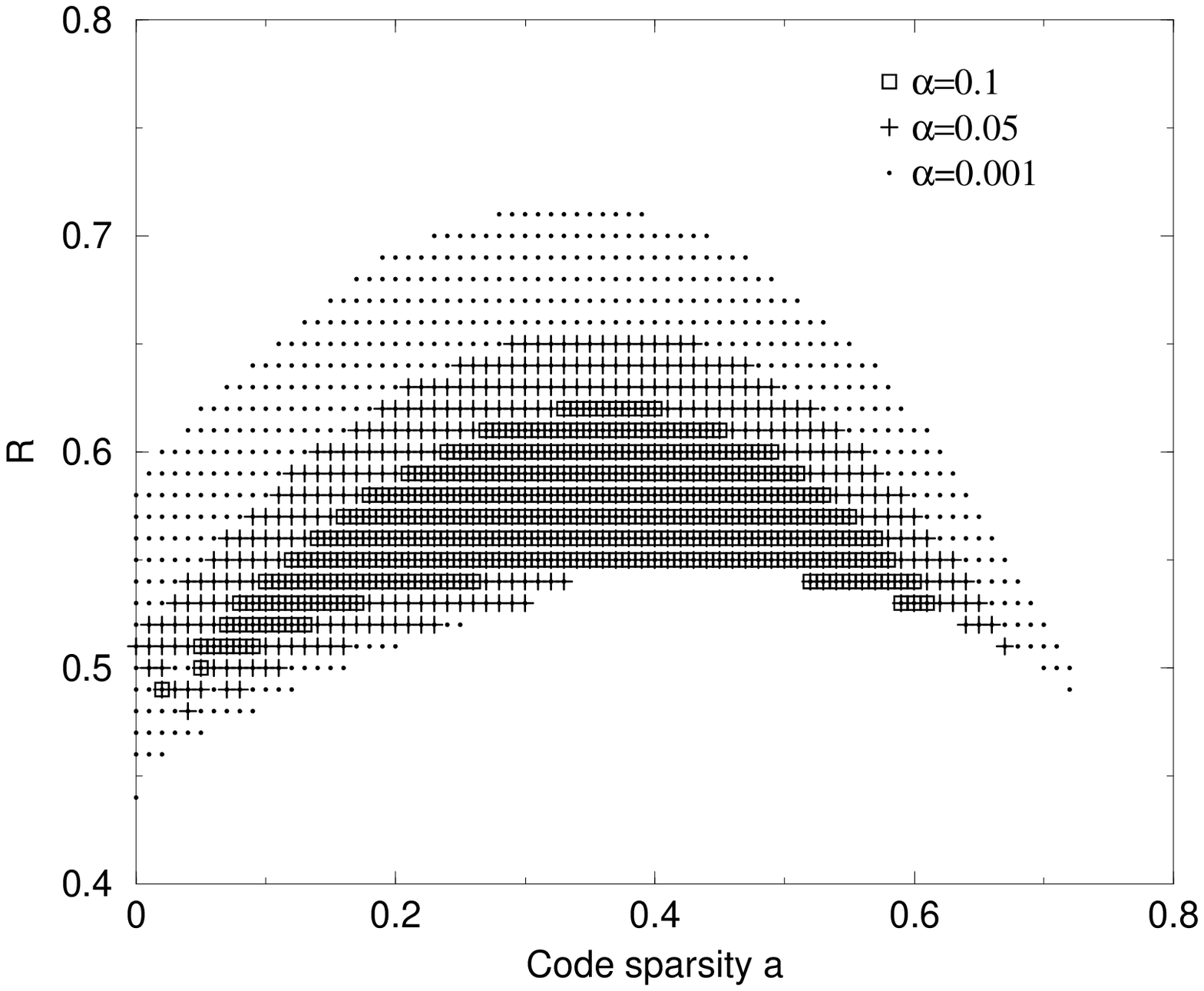}
\end{minipage}
\caption{\small
Left: Phase diagram R versus $a$ for $\alpha=0.001$ ($\alpha/c=0.02$). The SAS region, where local bumps are observed is relatively large. The Z region corresponds to trivial solutions for the overlap.
Right: SAS region for different values of $\alpha$. High values of $\alpha$ limit the SAS region. 
Note that not all of the values of $\alpha$ are feasible with arbitrary sparsity of the network $c$.
}
\label{phase}
\end{center}
\end{figure}

We have investigated the phase diagram when fixing the capacity taking into account the connectivity of the network. 
We have observed a stable area of solutions corresponding to the formation of local bumps, 
i.e. solutions with OP $m_1 \neq 0$ for wide range of the parameters load $\alpha$, sparsity $a$ and 
the retrieval asymmetry parameter $R$ (Fig.\ref{phase}). 
Note that $\alpha=p/N$ that is used in this paper,
although natural when working with eigenvectors, 
is slightly different from the usual one $\alpha'=P/(cN)=\alpha/c$.
The range of values of $\alpha/c$ corresponds to biologically plausible values.

According to this phase diagram, there exist states with $a=0$
and spatial asymmetric retrieval states. 
Therefore, the sparsity of the code
is not a necessary condition for SAS. 
Also it is clear that there is no necessity to use more complex model neuron
than a binary one in order to observe SAS.
It is worth mention that
the sparsity of the code makes SAS better pronounced.

On the other hand, there is no state
with SAS and $H_a=0$, and therefore the asymmetry between the
retrieval activity and the learned patterns is essential
for the observation of the phenomenon.

The diagram in Fig.\ref{phase} shows a clear phase transition with $R$.
For small values of $R$, the effect (SAS) is not present.
If $R>1$, then the only stable state is the trivial $Z$ state, as all the nonzero solutions are suppressed. 
Only for intermediate values of $R$, the bumpiness occurs. 
The load $\alpha$ of the network shrinks the area of parameters where bumps are observed. However, the area remains large enough for all values of $\alpha$.

Our observations show that the parameter $R$ is a very rigid one in the sense that 
the behavior of the system changes drastically for small changes of $R$. 
Richer behavior can be obtained when the term $H_a$ is replaced by a quadratic term of the total activity. 
Detailed analysis will be presented in a forthcoming publication \cite{EK1}.

\begin{figure}[t]
\begin{center}
\epsfxsize 5.7cm \epsfysize=5cm
\epsfbox{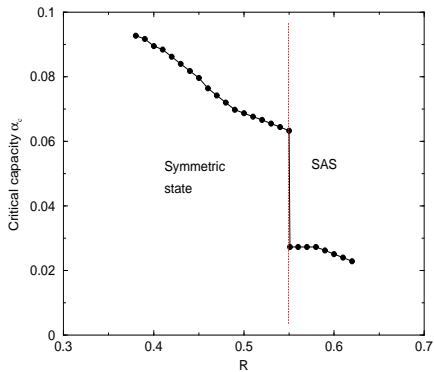}
\caption{\small
The critical storage capacity $\alpha_c$ as a function of $R$. $a=0.4$
The drop of $\alpha_c$ is clearly seen on the transition to SAS state.
}
\label{setcapacityR}
\end{center}
\end{figure}

When the network changes its behavior from homogeneous retrieval 
to spatial localized state, 
the critical storage capacity of the network drops
dramatically,
because effectively only the fraction of the network in the bump 
can be excited 
and the storage 
capacity drops proportionally to the size of the bump,
Fig.\ref{setcapacityR}.  As can be seen in this figure, in the phase transition point the capacity drops approximately twofold.

The local overlap of the bump remains strong but, 
in contrast to the linear threshold network, 
for binary neurons is always less than one.
However, if the threshold is determined as a function 
of the distribution of the 
internal neuronal fields, as in \cite{AleYasser}, 
the impact on the storage capacity of the binary network near the transition 
points is minor, as it can be seen by simulations.

These effects are interesting and deserves more attention in the future.

\section{Conclusion}
In this paper we have studied the conditions
for the appearance of spatial dependent activity in
a binary neural network model.

The analysis was done analytically,
which was possible due to the finite number of
relevant order parameters, 
and is compared with simulations.
The analytical approach gives a closed form for the equations
describing the different order parameters and
permits their analysis.

It was shown that the presence of the term $H_a$ is sufficient
for the existence of the SAS.
Nor asymmetry of the connection, neither sparsity of the code are
necessary to achieve spatial asymmetric states (bumps).
In our opinion, this is the main result of the paper.

The numerical solution of the analytical results,
as well as the simulations show that when
$H_a=0$
no SAS can be observed.
This is a good,
although not conclusive argument,
that the asymmetry between
the retrieval and the learning states might be a
necessary and not only sufficient condition
for the observation of spatial asymmetry states.

Detailed analysis of the problem will be presented
in a forthcoming publication.

\section*{Acknowledgments}
\vskip0.5cm
The authors thank A.Treves and Y.Roudi for stimulating discussions.

This
work is financial supported by the Abdus Salam Center for Theoretical Physics,
Trieste, Italy and by Spanish Grants CICyT, TIC 01-572,
DGI.M.CyT.BFM2001-291-C02-01 and the program ``Promoci\'on de la Investigaci\'on UNED'02''.

\end{document}